\documentclass[%
 reprint,
 amsmath,amssymb,
 aps,
pra,
]{revtex4-2}
\usepackage[table]{xcolor}

\usepackage{graphicx}
\usepackage{dcolumn}
\usepackage{bm}
\usepackage[colorlinks=true,linkcolor=blue,citecolor=blue]{hyperref}%
\usepackage{verbatim}

\begin{document}

\preprint{APS/123-QED}

\title{Probabilistic description of dissipative chaotic scattering}

\author{Lachlan Burton}
\author{Holger Dullin}%
\author{Eduardo G. Altmann}%
\affiliation{%
 School of Mathematics and Statistics, The University of Sydney, Australia
}%

\date{\today}

\begin{abstract}

We investigate the extent to which the probabilistic properties of a chaotic scattering system with dissipation can be understood from the properties of the dissipation-free system. For large energies $E$, a fully chaotic scattering leads to an exponential decay of the survival probability $P(t) \sim e^{-\kappa t}$ with an escape rate $\kappa$ that decreases with $E$. Dissipation $\gamma>0$ leads to the appearance of different finite-time regimes in $P(t)$. We show how these different regimes can be understood for small $\gamma\ll 1$ and $t\gg 1/\kappa_0$ from the effective escape rate $\kappa_\gamma(t)=\kappa_0(E(t))$ (including the non-hyperbolic regime) until the energy reaches a critical value $E_c$ at which no escape is possible.  More generally, we argue that for small dissipation $\gamma$ and long times $t$ the surviving trajectories in the dissipative system are distributed according to the conditionally invariant measure of the conservative system at the corresponding energy $E(t)<E(0)$. Quantitative predictions of our general theory are compared with numerical simulations in the Hénon–Heiles model.
\end{abstract}

\maketitle


\section{\label{sec:intro}Introduction}

\noindent Chaotic scattering is a prevalent phenomenon in open Hamiltonian systems which describes a wide range of physical phenomena, from complex astrophyiscal motions to the interactions of charged particles~\cite{Aguirre2001,NaglerJan2004,Tel2011}. In such problems, there is a bounded region of the potential, the \emph{scattering region}, in which generic trajectories remain confined ({\it survive}) for a long time before eventually escaping towards infinity. {\it Chaotic} scattering is well-understood as a phenomenon of \emph{transient chaos}~\cite{Tel2011} which explains how hyperbolic scattering leads to exponential decay of trajectories from the vicinity of a chaotic set and non-hyperbolic scattering leads to algebraic decay due to the stickiness around KAM islands~\cite{Ott02}. 

\par More recent studies have focused on the effect of perturbations on the general chaotic scattering picture, with the goal of expanding its applicability to other physical situations~\cite{Seoane2012}. For instance, it has been shown that the effect of white noise leads to enhanced trapping and intermediate-time regimes in the survival probability that depend on the noise strength~\cite{Seoane2012,Bodai2013}. Another important type of perturbation, on which we focus in this paper, is the effect of dissipative forces~\cite{Seoane2006,Seoane2012} which appear not only in mechanical systems but also in astrophysical problems (due to radiation pressure-like effects~\cite{Celletti2011}) and in advection of particles in fluids (inertial effects~\cite{Motter2002}).

\par Dissipation has a strong effect on chaotic scattering as it can lead to new asymptotic dynamics (e.g., attractors) that are not possible in the Hamiltonian setting and it necessarily alters the escape rate (or distribution of trajectory lifetimes) as monotonic energy decay due to dissipation implies that the average speed of any particle is decaying. It also non-trivially affects non-hyperbolic dynamics:  dissipation added to an area-preserving map had the effect of stabilising periodic orbits inside the KAM islands or destroying them entirely~\cite{Motter2002}; dissipation added to a mechanical system in which the energy decayed to a unique minimum~\cite{MONDRAGON1994} led to piecewise-adiabatic invariance of the action integrals, which manifested as complicated tube-like structures in the extended phase space. In any case, the destruction of the KAM islands, the phase space objects responsible for stickiness~\cite{Ott02}, causes the decay law of the survival probability to become exponential. The fractal properties of scattering systems is also strongly affected by dissipation, with the appearance of different regimes at different strengths of the dissipation~\cite{Motter2003,Seoane2006,Seoane2012}. 

\par A novel theory to describe the motion of trajectories in non-linear systems with dissipation is to view it as \emph{doubly-transient chaos}~\cite{MotterAdilsonE.2013DtcG,Tel2011,Vilela_2021}. As suggested by its name, the idea is to view the invariant sets underlying transient chaos to be transients due to the loss of energy. This theory has been shown to describe the systems in which the dissipation acts as a non-autonomous element in the dynamics, with mechanical systems like the magnetic pendulum receiving particular attention~\cite{MotterAdilsonE.2013DtcG}. The transient chaos properties of such systems are found to be surprisingly distinct from those in the conservative case. Key features include time-dependent settling rates to attractors, and scale-dependent fractal dimensions of attractor basin boundaries. Ref.~\cite{Karolyi_2021} explored further properties of doubly-transient chaos by analysing the time evolution under dissipation of phase space structures of the dissipation-free system, proposing new tools to analyse how the complexity of the motion decays in time. 

Most existing works on doubly-transient chaos have focused on closed mechanical systems~\cite{MotterAdilsonE.2013DtcG,Karolyi_2021,Chen2017}, wherein every trajectory of an otherwise permanently chaotic system eventually settles to a fixed point when dissipation is added. It is not clear from these studies what to expect when dissipation is added to open systems whose generic behaviour is already transient. Scattering systems are a natural example, as the only orbits with infinite lifetimes can often form measure-zero fractal sets. An interesting exception is Ref.~\cite{Vilela_2021}, which applies the theory of doubly-transient chaos to a Hamiltonian system subject to parameter drift whose unperturbed dynamics is transiently chaotic (a decaying open flow), describing the escape process using time-dependent (effective) escape rates that can be understood as the decaying system evolving along snapshots of the system without parameter drift. An important simplifying property of the system studied in that work is that all trajectories experience the same dynamics at time $t$. In more typical scenarios, as the ones we consider in this paper, the dissipation depends on the trajectory (e.g., on the velocity) and thus different trajectories experience a different unperturbed dynamics (have a different energy) at the same time $t$. 

\par In this paper we build on the previous works on dissipative scattering and doubly transient chaos to propose a probabilistic description of the effect of dissipation on scattering problems. Instead of focusing on invariant sets and specific dynamics quantities (e.g., escape rates or fractal dimensions), we focus on the probability of a trajectory being in a phase space region at long-times $t$ and show how this allow us to derive the observable quantities of interest, such as the survival probability as a fucntion of time $P(t)$ and the phase-space (fractal) distribution of long-living trajectories. In particular, we propose that the trajectories at time $t$ and energy $E(t)$ will be distributed according to the conditionally invariant measure (c-measure) of the Hamiltonian system with the same (fixed) energy. Importantly, we specify the ranges of time $t$, energy $E$, and dissipation strength $\gamma$ for which this holds, showing that this description can be made arbitrarily precise even at finite (yet small) dissipation $\gamma$. This description has important and testable consequences. Namely, dynamical properties of ensembles of trajectories evolving on the dissipative system can be approximated by the properties of the dissipation-free system with the corresponding energy, but differences are expected at small spatial scales. To account for these differences, we describe an entire ensemble at time $t$ as a combination of c-measures over the range of trajectory energies, weighted by the distribution of trajectories at these energies. We also show that dissipation typically acts to slow down the escape process, leading to enhanced trapping that can be well-approximated from the relevant quantities in the Hamiltonian system. This works particularly well for small dissipation, large initial energy,  and if the underlying conservative dynamics is hyperbolic.

\par We divide the paper as follows. We start in Sec.~\ref{sec:conservative} by discussing the well-known case of conservative (no dissipation) chaotic scattering. We briefly discuss how the survival probability $P(t)$, escape rate $\kappa$, and conditionally-invariant measure $\mu_c$ are related and depend on $E$. We also introduce the Hénon–Heiles system and present some visualisations of the aforementioned quantities in the conservative setting.  In Sec.~\ref{sec:dissipative} we describe how we add dissipation to the Hénon–Heiles system, and show how the survival probability and fractal properties change. We go on to introduce our probabilistic model for the dissipative system and test our ideas by comparison with numerical results. General discussions and conclusions appear in Sec.~\ref{sec:conclusion} while details of the numerics appear in the Appendix~\ref{app:numerics} and in the repository~\cite{Repo}.

\section{Conservative Scattering}\label{sec:conservative}

\subsection{Theory}
\subsubsection{Setting}
We are interested in chaotic scattering defined as the dynamics of Hamiltonian systems for which trajectories approach arbitrary large positions in configuration space after performing a transiently chaotic motion ~\cite{Lau1991}. The simplest setting in which this happens are Hamiltonian systems in two dimensions $(x,y)$ with energy for a unit mass given by
\begin{equation}\label{eq:H}
    H = \frac{1}{2}(p_x^2 + p_y^2) + V(x,y),
\end{equation}
where $p_x,p_y$ are the momenta and $V(x,y)$ is a potential function that leads to a non-linear (non-integrable) dynamics in a trapping region $\Omega$ close to the origin. In this setting, the energy $E = H$ is preserved and acts as a control parameter. For small energies $E$, the trajectories typically remain trapped close to the origin performing regular (periodic or quasi-periodic) motion and scattering is not possible. Increasing the energy, one typically observes a regime in which regular and (transiently) chaotic motion coexists, while for very large energies most trajectories quickly leave the region of interest apart from a cantor set of orbits that remain. 

To study the continuous-time flow induced by the Hamiltonian~(\ref{eq:H}), we consider its intersection with a Poincaré surface of section. This allow us to visualise the trajectories in a 2D space -- for example the configuration space $(x,y)$ -- and it defines a true-time map, i.e., a discrete-time system $h$ mapping consecutive intersections with the section in which the time $T$ between intersections is recorded~\cite{Altmann2013,RevModPhys.85.869}. We start trajectories in the surface $(x,y) \in \Omega$ at the same energy $E$. This choice of initial conditions inside the scattering region differs from usual scattering settings, in which particles approach $\Omega$ but start far from it. Our choice simplifies numerical computations and is more suitable for settings in which dissipation acts globally  (for instance, orbits coming from infinity would never reach $\Omega$). In the conservative setting, this choice affects the stickiness exponent in the intermediate-$E$ (non-hyperbolic) regime~\cite{PIKOVSKYAS1992}. However, it has no significant impact on our simulations (which are typically at larger $E$) and conclusions.

\subsubsection{Survival Probability}\label{sssec:survival}

Transient chaos theory provides a description of the transient dynamics using the properties of the non-attracting fractal set of orbits that remain in the system for $t\rightarrow \pm \infty$. This \emph{chaotic saddle} and its invariant manifolds control the rate at which nearby trajectories leave the system and thus the long-term properties of the survival probability $P(t)$ of typical initial conditions~\cite{Tel2011}. More precisely, we define $P(t)$ as the proportion of an initial ensemble of trajectories that remain in the system (have not 
 \emph{escaped}) until time $t$. Equivalently, $P(t)$ is the probability that a trajectory remains inside $\Omega$ after time $t$. For uniformly hyperbolic chaotic saddles, we observe
\begin{equation} \label{eqn:kappa_hyperbolic}
    P(t) = 
    \begin{cases}
        \text{irregular} & t \in [0,\tau_0]\\
        f_0 e^{-\kappa t} & t > \tau_0,
    \end{cases}
\end{equation}
where the initial period $[0,\tau_0]$ depends sensitively on the choice of initial density e.g., some trajectories may escape quickly before approaching the saddle, and the exponent $\kappa$ is called the \emph{escape rate}, and characterises the global (in)stability properties of the saddle.

\par For systems with mixed phase space, such as Hamiltonian systems in parameter regimes allowing for KAM tori, the escape dynamics is more complicated~\cite{Ott02,PIKOVSKYAS1992}. A chaotic trajectory that comes too close to a KAM island can spend an arbitrarily long time in its vicinity before leaving. This {\it stickiness} phenomenon happens because the local Lyapunov exponents approach $0$ near such regions. As a consequence, such orbits will exhibit an effective escape rate much lower than those that never come close to these regions. The effect on the survival probability is that, on some timescale $\tau^{\star}$, the dominant asymptotic decay switches from exponential to a power law~\cite{Tel2011,RevModPhys.85.869}:

\begin{equation} \label{eqn:kappa_nonhyperbolic}
    P(t) \sim 
    \begin{cases}
        \text{irregular} & t \in [0,\tau_0]\\
        f_0 e^{-\kappa t} & \tau_0 < t < \tau^{\star},\\
        f_{\star} t^{-z} & t > \tau^{\star},
    \end{cases}
\end{equation}
where the time $\tau^\star$ indicates the transition from exponential to algebraic decay due to stickiness.
In such cases, one can speak of the saddle as possessing both hyperbolic and non-hyperbolic components~\cite{Motter2003,RevModPhys.85.869,Altmann2009}. In the system we consider in this paper, the phase space of the dissipation-free system can be either strongly chaotic or mixed, depending on the value of the energy $E$.

\subsubsection{Conditionally-invariant measures}\label{sssec:cmeasure}
For long times, the probability of a surviving trajectory being in any region $A \in \Omega$ approaches $\mu_c(A)$, where $\mu_c$ is the conditionally-invariant measure of the system and is distributed along the unstable manifold of the saddle \cite{RevModPhys.85.869,DemersMarkF2006Erac,Tel2011}. This measure acts as an attractor for generic ensembles of initial conditions (which intersect the stable manifold of the saddle) so that a smooth ensemble of points initiated will after some time $t^* \gg \frac{1}{\kappa}$ be effectively distributed according to $\mu_c$. It is $\mu_c$ which characterizes the process by which points leave $\Omega$ and it describes the relevant quantities in chaotic scattering. For instance,  the escape rate $\kappa$ of a hyperbolic discrete-time map $h$ is given by~\cite{PianigianiYorke1979}

\begin{equation}\label{eqn:kappa_cmeasure_discrete}
    \kappa = -\log (1-\mu_c(\mathcal{E})),
\end{equation}
where $\mu_c(\mathcal{E}) = 1- \mu_c(h^{-1}(\Omega))$ is the c-measure of the set $\mathcal{E}$ of trajectories that exit $\Omega$ in one iteration of the map $h$.
For true time maps, as considered here, $\kappa$ is given implicitly by \cite{Altmann2013}
\begin{equation}\label{eqn:kappa_cmeasure_truetime}
    \int_\Omega e^{\kappa T} d\mu_c = 1 + \mu_c(\mathcal{E}),
\end{equation}
where $T$ is the time between intersections of the Poincar\'e surface of section.

\begin{figure*}
    \centering
    \includegraphics[width=0.9\linewidth]{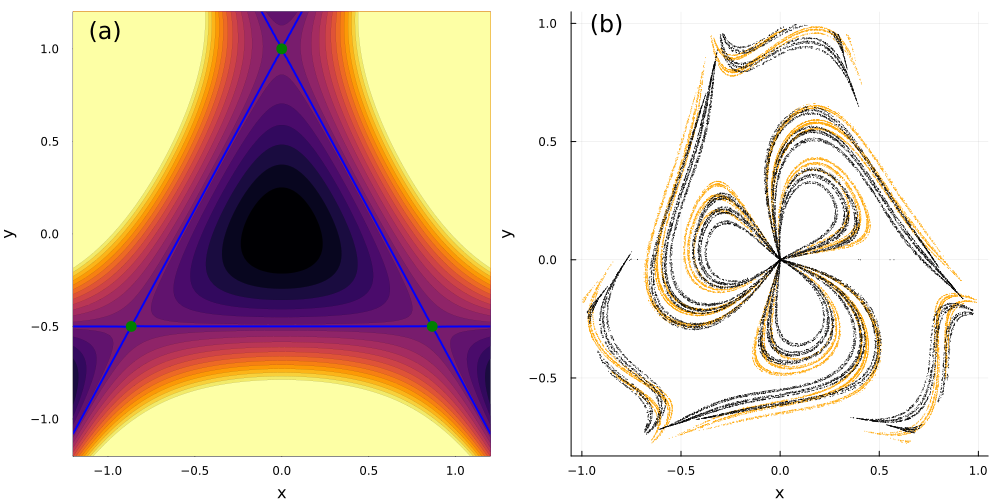}
    \caption{Phase-space section $(x,y)$ of the Hénon–Heiles system~(\ref{eqn:hh_ham}). \textbf{(a)}: Contours in the  potential~(\ref{eqn:hh_potential}) for 20 values of the energy $E \in (0.0,0.5)$. The contour corresponding to $E = \frac{1}{6}$ is shown in blue. The saddle points of the potential, are indicated by green dots. Darker colours correspond to smaller values of $E$. \textbf{(b)}: Support of the conditionally-invariant measure in the Hénon–Heiles system for $E = 0.35$ (orange points) and $E = 0.3$ (black points). These approximations were computed via the sprinkler method \cite{Tel2011}: integrating a large uniform ensemble of initial conditions for a long time $t \gg \frac{1}{\kappa}$ and plotting the positions of the points still inside $\Omega$ at the next intersection with the Poincare surface. }
    \label{fig:contour_two_manifolds}
\end{figure*}

\subsection{Numerical Results}
\subsubsection{\label{sec:modelsys}Model System}

We study trajectories of the Hénon–Heiles system\cite{HenonHeiles1964}, a Hamiltonian system introduced as a toy of model of stellar motion in a 2D galactic potential that shows the generic properties discussed above. 
The potential function in Eq.~\eqref{eq:H} is
\begin{equation}
\label{eqn:hh_potential}
    V(x,y) = \frac{\omega^2}{2}(x^2 + y^2) + \lambda\left(x^2y - \frac{1}{3}y^3\right),
\end{equation}
which combines a harmonic potential with cubic terms that create non-integrable dynamics with parameters $\omega$ and $\lambda$. 
Through Hamilton's equations, we are led to a four dimensional ODE:
\begin{subequations}\label{eqn:hh_ham}  
\begin{align}
    &\dot{x} = \frac{\partial H}{\partial p_x}  = p_x,\\
    &\dot{y} = \frac{\partial H}{\partial p_y} = p_y,\\
    &\dot{p_x} = -\frac{\partial H}{\partial x} = -\omega^2x - 2\lambda xy,\\
    &\dot{p_y} = -\frac{\partial H}{\partial y} = -\omega^2y - \lambda(x^2 - y^2).
\end{align}
\end{subequations}

The system is often studied in a perturbative setting (i.e, for $\lambda \ll 1$) to demonstrate how chaos emerges as the harmonic terms in the potential become less dominant. However, here we fix ourselves firmly in the non-integrable setting, taking $\lambda = \omega = 1$. See Appendix \ref{app:numerics} for details of the numerical simulations, including how initial conditions were chosen, the Poincaré section, the criteria for escape/non-escape of trajectories from $\Omega$.

\subsubsection{Scattering dynamics}
As is typical for these kinds of systems, there is a certain value of the Hamiltonian, which we call the \emph{critical energy} and denote by $E_c$, below which the contours of the potential are closed and escape from near the origin becomes impossible. For the potential~(\ref{eqn:hh_potential}) with $\omega=\lambda=1$, $E_c=1/6$. For $E>E_c$, three openings are created and particles may escape through any of the \emph{exit channels}. We emphasize that it is only for energies above $E_c$ that escape is possible, and thus transient chaos is observable.  At $E = E_c$, the potential contours form an equilateral triangle with vertices at saddle points. This is shown graphically in Fig.~\ref{fig:contour_two_manifolds}(a). We present an approximation of the support of the conditionally-invariant measure of the Hénon–Heiles system (in the same Poincaré surface of section as the initial conditions, see App.~\ref{app:initial_conditions}) in Fig.~\ref{fig:contour_two_manifolds}(b). The measure is plotted for two values of the energy as a visual comparison of how an initial ensemble relaxes to a different object at each energy.

\begin{figure}
    \centering
    \includegraphics[width=0.5\textwidth,height=7cm]{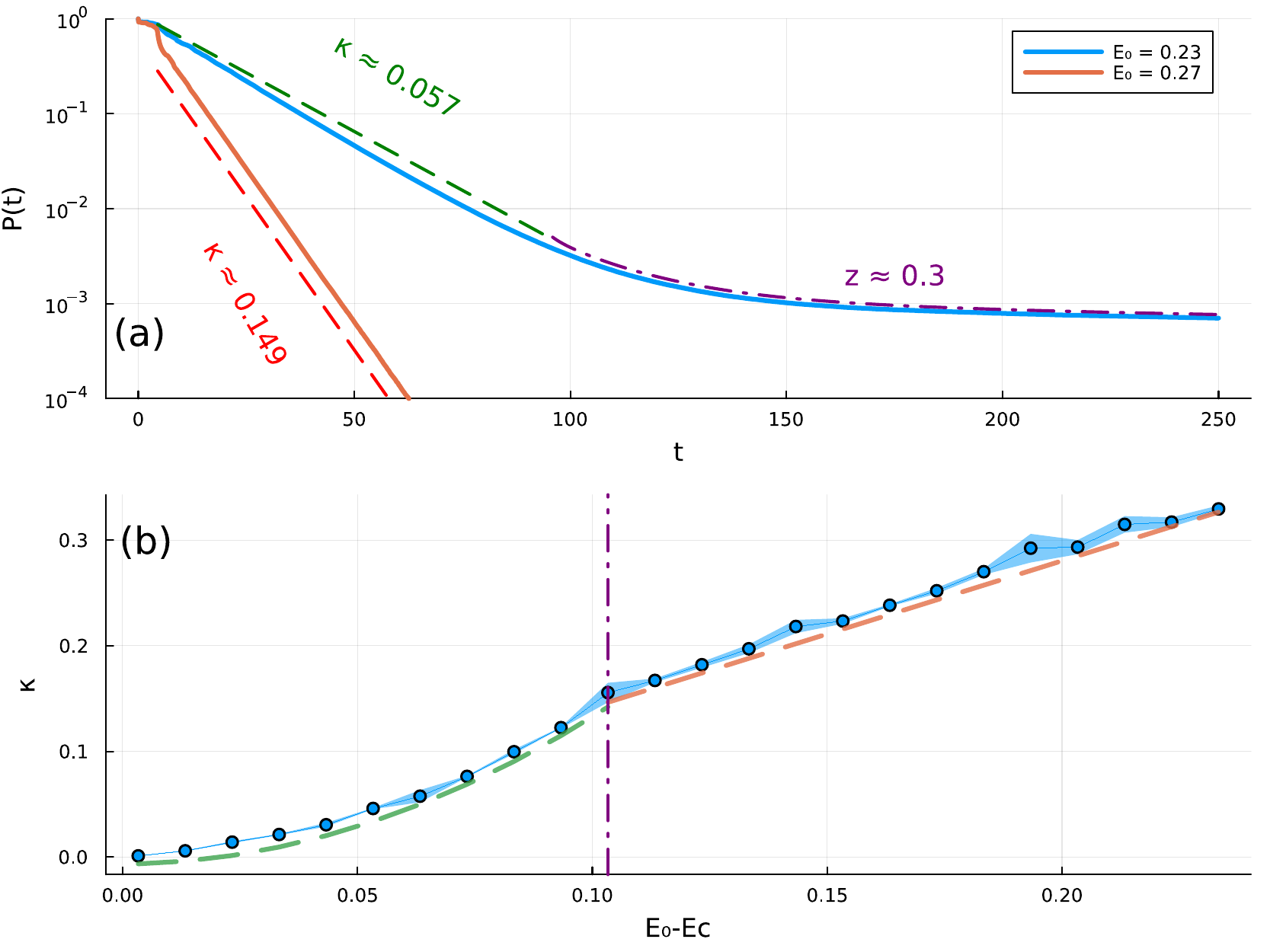}
    \caption{Survival probability $P(t)$ and escape rate $\kappa=\kappa_0$ in the Hamiltonian system. \textbf{(a)}: $P(t)$ for 2 values of the energy $E_0 > E_c = 1/6$ (see legend).  For the smaller energy ($E_0 = 0.23$), a crossover occurs between an exponential and a power law decay at at $t=\tau^\star \approx 100$, as described in Eq.~\eqref{eqn:kappa_nonhyperbolic}. For the larger energy ($E_0=0.27$), there is no crossover and the decay is exponential as in Eq.~\eqref{eqn:kappa_hyperbolic}, which suggests a uniformly hyperbolic dynamics. The dashed lines have been added as a visual aid, with the approximate decay exponents for the exponential and power law regimes annotated. $N_0 = 10^7$ initial conditions were used for each curve.  \textbf{(b)}: Dependence of the escape rate $\kappa_0$ on the energy $E_0$ in the Hamiltonian system. Note that the x-axis is translated by $E_c$. Each point is the mean of 4 separate estimates of $\kappa$ made by least-square fits of the logarithm of $P(t)$ across different time intervals. There is an observed non-linear scaling which ends at around $E_0= 0.27$ (marked with the purple dash-dotted line) after which the dependence becomes linear. Parametric fits of the non-linear (green dashes, $\kappa = 13.63(E-E_c)^2 + 0.03(E-E_c)$) and linear (orange dashes, $\kappa = 1.41(E-E_c)$) regimes are overlaid. }
    \label{fig:kappa-E-dependence}
\end{figure}

The escape dynamics in the Hamiltonian problem have been extensively studied~\cite{Aguirre2001, Zotos2017}. A typical orbit moves about the scattering region until it passes through one of the three exit channels and escapes to infinity. The existence of special unstable periodic orbits, called \emph{Lyapunov orbits}~\cite{1990A&A...231...41C}, allow for the definition of a meaningful criterion for escape from the region around the origin. These orbits straddle the exit channel openings, and persist for all energies $E > E_c$. Trajectories which cross these orbits with outward velocity vectors can never return to the central region. For our problem, we thus define the scattering region $\Omega$ to be the region of configuration space bounded by the potential contours $V = E$ and the Lyapunov orbits at energy $E$ connecting neighbouring contours.

\subsubsection{Survival Probability}

Numerical computations of the survival probability in system~\eqref{eqn:hh_potential} are summarized in Fig.~\ref{fig:kappa-E-dependence}. The top panel confirms the existence of all the generic regimes described in Sec.~\ref{sssec:survival}, depending on the initial energy $E$. Importantly, even if for small energies the existence of KAM islands hinders the decay to be non-exponential, there is an effective escape rate $\kappa$ at intermediate times $t$~\cite{Motter2003,RevModPhys.85.869} -- as described in Eq.~\eqref{eqn:kappa_nonhyperbolic} - that governs the escape of the majority of the initial conditions. It is thus possible to investigate the dependency of the (effective) escape rate $\kappa$ on the energy $E$ for all values of $E > E_c$. The numerical results in Fig.~\ref{fig:kappa-E-dependence}(b) show a continuous dependence of $\kappa$ on $E_0$, with a roughly linear dependency for $E$ in the hyperbolic regime.

\section{Dissipative scattering} \label{sec:dissipative}

\subsection{Numerical Results}
\subsubsection{Dissipation}

We study the effects of weak dissipation on the escape dynamics of the open system by introducing an 
additional term to Eqs.\eqref{eqn:hh_ham} modelling a drag linearly dependent on the velocity $v$. The dynamics of the resulting dissipative system is given by

\begin{subequations}
\begin{align}
    &\dot{x} = p_x,\\
    &\dot{y} = p_y,\\
    &\dot{p_x} = -\omega^2x - 2\lambda xy - \gamma p_x,\\
    &\dot{p_y} = -\omega^2y - \lambda(x^2 - y^2) - \gamma p_y,
\end{align}
\end{subequations}
where $\gamma\ge0$ controls the strength of the dissipation. The system is no longer Hamiltonian and the energy of a trajectory decays as
\begin{equation} \label{eqn:energy_diffeq}
   \begin{aligned}
    \frac{dE}{dt} &= -\gamma (p_x^2 + p_y^2) = -\gamma v^2,\\
    & = -2\gamma (E - V(x,y)). 
    \end{aligned} 
\end{equation}
The decay rate of the energy will thus differ for each initial condition depending on the potential $V(x,y)$ they experience.  For large times $t$ long-living trajectories perform a similar movement in $\Omega$. The variations across such trajectories will be small and the potential can be approximated (on average across $t$) as a fraction of the total energy $V(x,y) = \frac{\mu}{2} E$, where $\mu \approx 1$ is a proportionality factor. Using this approximation in Eq.~\eqref{eqn:energy_diffeq}, we obtain 

\begin{equation}
    E(t) = E_0e^{-\mu\gamma t}.
    \label{eqn:energy_exp_decay}
\end{equation}
For damped oscillations around a stable fixed-point, the linearised system has eigenvalues with real part $-\gamma/2$, and thus $E$ should decay at a rate $\gamma$ and thus $\mu=1$.

\begin{figure}
    \centering
    \includegraphics[width=0.9\linewidth]{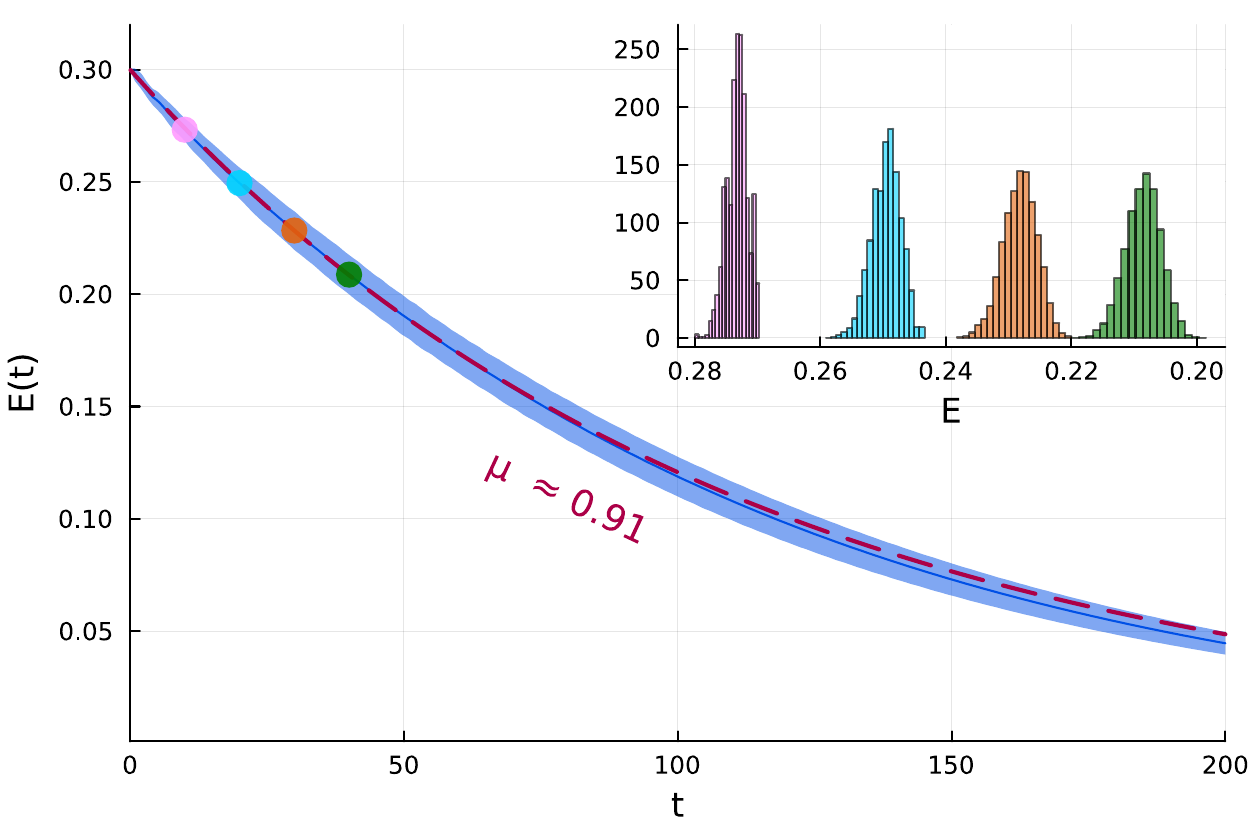}
    \caption{Energy dependence on time in the dissipative system. The dark blue line shows the mean energy $\langle E\rangle$ computed over an ensemble of trajectories inside $\Omega$ at time $t$. The blue ribbon shows 3 standard deviations. The dashed red line is a fit to Eq.~\eqref{eqn:energy_exp_decay} obtained via least-squares of the log-transformed data. \textbf{(Inset)}: Histograms showing the distribution of energies at times corresponding to the coloured dots in the main plot. The parameters are $(E_0,\gamma) = (0.3,0.01)$.}
    \label{fig:energy_decay}
\end{figure}

We numerically test the validity of Eq.~\eqref{eqn:energy_exp_decay} for our system in Fig.~\ref{fig:energy_decay}. The decay is well-described by an exponent of $\mu = 0.91$ for the time interval $[0,200]$, after which the only remaining orbits will be slowly collapsing towards the origin. The energy of such long-lived orbits decays faster than the overall fit (the red-curve slightly overestimates the numerical results for large $t$). This, and the expectation from the calculations of the linearized system mentioned above, suggest that $\mu$ approaches $1$ for increasing $t$.

\subsubsection{Survival Probability}

We now focus on the survival probability $P(t)$ of trajectories. Dissipation creates an attractor at the origin corresponding to a point of zero energy. The basin of attraction of this fixed-point attractor is comprised by initial conditions in or around  KAM islands~\cite{Motter2002,Celletti2011} and orbits initialised near the stable manifold of the chaotic saddle of the conservative system. These are long-living trajectories for which dissipation acts for sufficiently long time to bring their energy to $E_c$.
In comparison to the conservative case, the attractor provides an additional way for an orbit to ``leave" the system. We consider the lifetime or {\it settling time} of such orbits, for purposes of measuring $P(t)$, to be 
the time at which $E(t) = E_c$ because for $E(t) < E_c$ escape becomes impossible (see App.~\ref{app:escape_condition} for further details).

\begin{figure}
    \centering
    \includegraphics[width=0.9\linewidth]{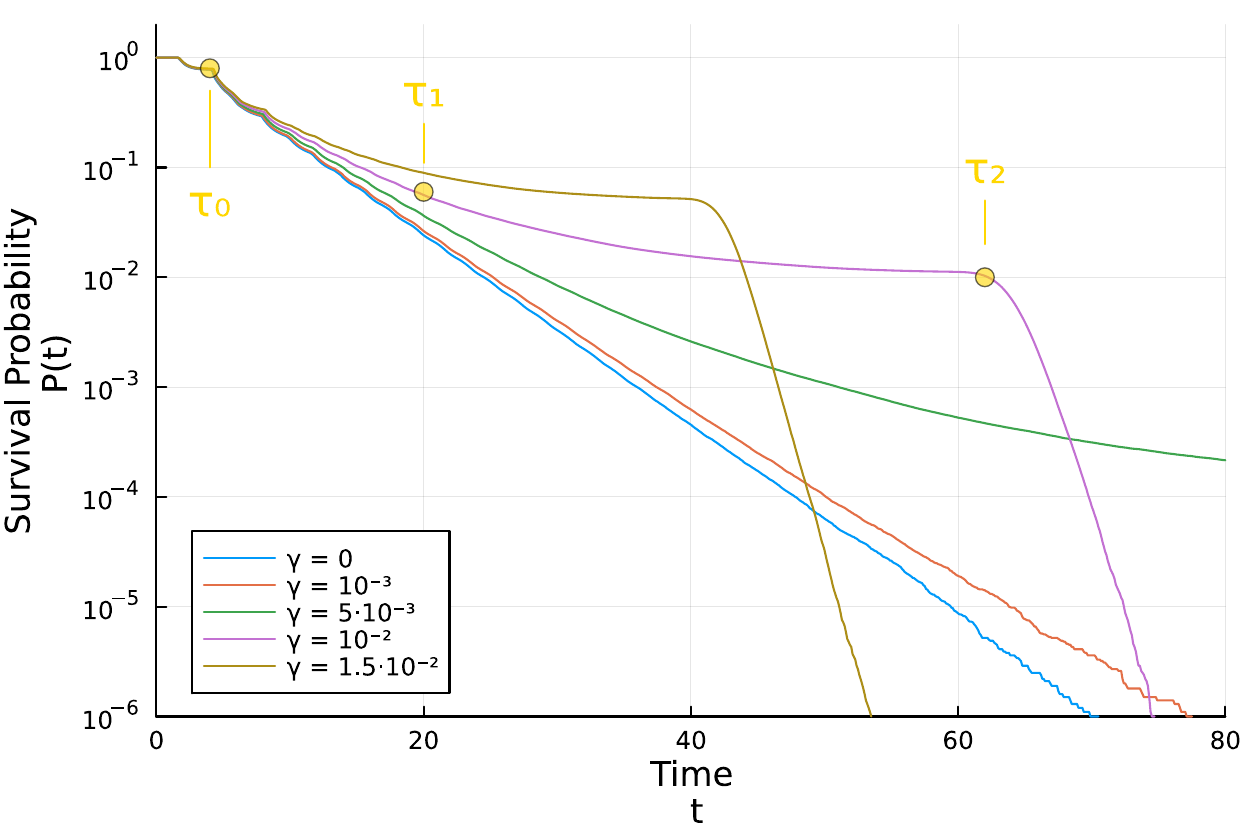}
    \caption{Distribution of trajectory lifetimes in the dissipative Hénon–Heiles system for various small dissipations $\gamma$. The survival probability for the Hamiltonian setting ($\gamma = 0$, blue curve) is added for comparison. For the purple curve $(\gamma = 10^{-2})$, the approximate positions of the timepoints $\tau_0,\tau_1$ and $\tau_2$ are marked.}
    \label{fig:survprob}
\end{figure}

The numerical results for $P(t)$ for various dissipation strengths $\gamma$ are shown in Fig.~\ref{fig:survprob}. Different decay regimes and the effects of increasing $\gamma$ are readily visible. After an initial period, the $P(t)$ curve deviates from the simple exponential of the Hamiltonian system (blue line), showing a slower decay (increased trapping). The time $\tau_1$ that marks the start of the enhanced trapping regime can be thus defined as
\begin{equation}\label{eqn:tau1}
    \tau_1 := \min \{t\ge\tau_0 | P_{\gamma > 0}(t) = \alpha P_{\gamma = 0}(t)\},
\end{equation} 
where $\alpha > 1$ is an arbitrary constant that demarcates when significant deviations from the non-dissipative system are observable (we use $\alpha = 2$ for our experiments). When $P(t)$ is almost constant (no escape), there is a sudden and dramatic increase in the decay that continues until there are no points left. We denote by $\tau_2$ the time that marks the end of the enhanced trapping regime, when $P(t)$ decays quickly (exponentially) towards zero. The numerical results show that both $\tau_1$ and $\tau_2$ increase with decreasing $\gamma$, consistent with the expectation that $P(t)$ varies continuously with $\gamma$.  Next we show how these numerical observations can be theoretically explained.

In order to characterize the observed survival probability $P(t)$, we introduce three characteristic times $\tau_0,\tau_1$, and $\tau_2$. The interval $[0,\tau_0]$ is the period during which the decay is highly-dependent on the choice of initial ensemble (see Sec.~\ref{sec:conservative}). Consequently, $\tau_0$ is effectively independent of $\gamma$.

\begin{figure*}[bt]
    \centering
    \includegraphics[width = 0.9\linewidth]{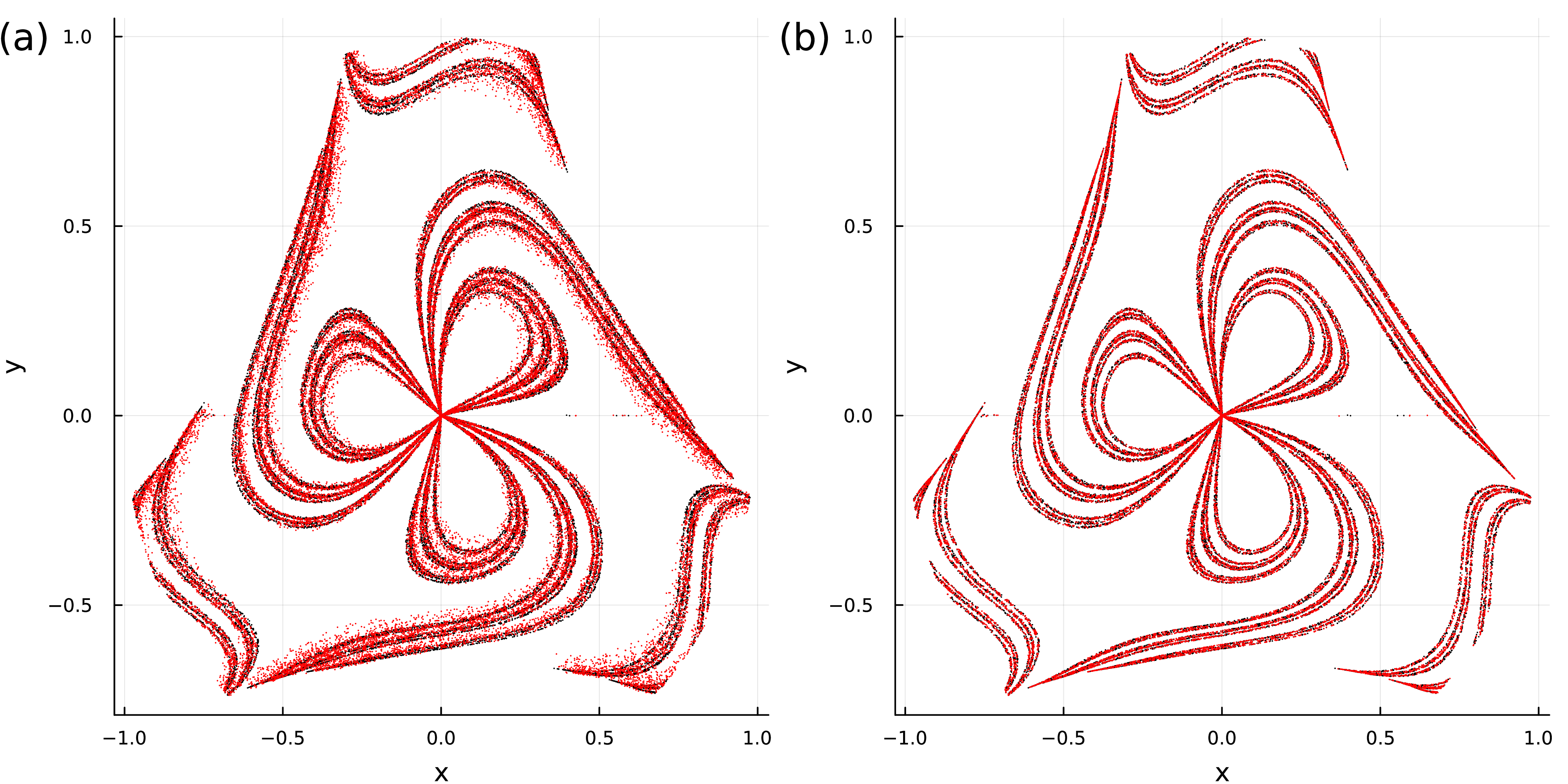}
    \caption{Comparison of the c-measure of the Hamiltonian system with $E = 0.3$ (black points) with the distribution of trajectories in the dissipative system $(E = 0.35, \gamma = 0.01)$ at $t=t^{\star}\approx 17 \gg \tau_0$ (red points.) \textbf{(a)}: all points that survived until $t^{\star}$ (the first intersection for $t\ge t^\star$ with the surface of section is shown). \textbf{(b)}: only the surviving points whose energy at time of crossing $t\ge t^\star$ was in the interval $[0.295,0.305]$, that is, within $1.67\%$ of $E_1$.$N = 10^7$ initial conditions were used and approximately $3\cdot10^4$ are plotted in each case. }
    \label{fig:um_comparison}
\end{figure*}

\subsection{Theoretical model}
\label{Sec:Model}
\subsubsection{Probabilistic description}
We now propose a probabilistic description to understand dissipative scattering systems, which will be compared to the numerical results reported above.  Dissipation leads to a trajectory-dependent decay of the energy $E(t)$ and the scattering process ends when the critical energy is met $E(t)=E_c$. The expected time at which this happens, which we associate to the transition time $\tau_2$ introduced above, is obtained solving Eq.~\eqref{eqn:energy_exp_decay} as
\begin{equation}\label{eq:tau2}
\tau_2 = \frac{1}{\mu \gamma}\log \left( \frac{E_0}{E_c} \right).
\end{equation}

The interesting dynamics happens at large times ($t>\tau_1$, when effects of dissipation become relevant) smaller than $\tau_2$. The key idea is to use the properties of the non-dissipative system to describe the scattering in this regime. We thus assume the unperturbed system to be fully chaotic and to have a well defined ($E$-dependent) conditionally-invariant measure $\mu^E_c$ in a common scattering region $\Omega$ for a wide range of (large) energies $E$ (i.e., $\int_\Omega d\mu^E_c =1$ for any $E\gg E_c$). The challenge here is that the c-measure is an attractor for surviving trajectories and thus provides a precise description only in the limit $t\mapsto \infty$, which is not applicable in our case since scattering happens only for $t<\tau_2$ (i.e., for $t>\tau_2$ there is no scattering or invariant set). 
\ 
Inspired by the idea of snapshot saddles and pullback attractors~\cite{Tel2011}, we focus on surviving trajectories that at a given large time $t$ have energy $E=E(t)$ and that started at arbitrarily early times with arbitrarily large energies (i.e., in a limiting case, we take $t_0 \mapsto -\infty$ and $E\mapsto +\infty$). We posit that, for $ \tau_0 \ll t<\tau_2$, the probability of finding such a surviving trajectory in a region of $\Omega$ is increasingly well described (for $t \lessapprox \tau_2 \rightarrow \infty$ and $\gamma \rightarrow 0$) by the c-measure of the conservative system $\mu_c^E$ with $E=E(t)$. From Eq.~(\ref{eq:tau2}), $\tau_2$ and thus the time the trajectory has to relax to $\mu_c$ can be made arbitrarily large by reducing $\gamma$ or increasing $E_0$. This would suggest that the description can be made arbitrarily precise for any fixed $\gamma$ or $E_0$. However, for $\gamma$ too large ($\gamma \gtrapprox 1$) the dynamics of the dissipative system are no longer comparable to the non-dissipative system, and so we restrict to a small $\gamma$ regime.
 
We now consider the consequences of this to an ensemble of trajectories, as investigated in our numerical experiments. 
After some sufficient time $t > \tau_0$, the energy $E(t)$ of trajectories that have not yet escaped will be distributed within some range $0 \le E(t)_{\text{min}} \le E(t) \le E(t)_{\text{max}} < \infty$ with a probability density $\rho(E = E(t))$ (which depends on the initial ensemble). The probability of finding a surviving trajectory at time $t$ in any region $A \subset \Omega$ is given by the composition of the sub-ensembles of trajectories at a specific energy $E = E(t)$ as
\begin{equation} \label{eqn:c-measure_model}
\mu_t(A) = \int_{E(t)_{\text{min}}} ^{E(t)_{\text{max}}} \mu_c ^E (A)\rho(E) dE.
\end{equation}
The convergence of surviving trajectories towards Eq.~(\ref{eqn:c-measure_model}) depends on the time $t$ available to relax towards $\mu_c^E$, which can be made arbitrarily large varying the initial energy $E_0$ and escape rate $\gamma$ as described in Eq.~(\ref{eq:tau2}). 
While this description assumes the existence of $\mu_c$, it is agnostic to the specific dynamics of the system and the form of dissipation considered. For instance, if we consider the case of a discrete Hamiltonian system with parameter drift as in Ref.~\cite{Vilela_2021}, then all trajectories at time $t$ have the same energy $(E(t)_\text{min} = E(t)_\text{max} = E(t))$ and so the density $\rho$ reduces to a Dirac delta function
\begin{equation}
\rho(E) = \delta(E-E(t)),
\end{equation}
and Eq.~\eqref{eqn:c-measure_model} yields
\begin{equation}
    \mu_t(A) = \int_{E(t)_\text{min}}^{E(t)_\text{max}} \mu_c ^E(A) \delta(E-E(t)) dE = \mu_c ^{E(t)}(A).
\end{equation}
Given the connection between the c-measure and the escape rate $\kappa$ in Eq.~(\ref{eqn:kappa_cmeasure_truetime}), we see that our description recovers as a particular case the finding made in Ref.\cite{Vilela_2021} about the instantaneous escape rate mirroring that of the Hamiltonian system at the new parameters. We may also account, through the energy density $\rho(E(t))$, for situations where the initial energy is not the same for each orbit, a common scenario in systems where particles are released from rest inside a potential. Generally speaking, the more information one has about $\rho$, the better one can approximate the state of the dissipative system.

\begin{figure*}[bt]
    \centering
    \includegraphics[width=0.48\linewidth]{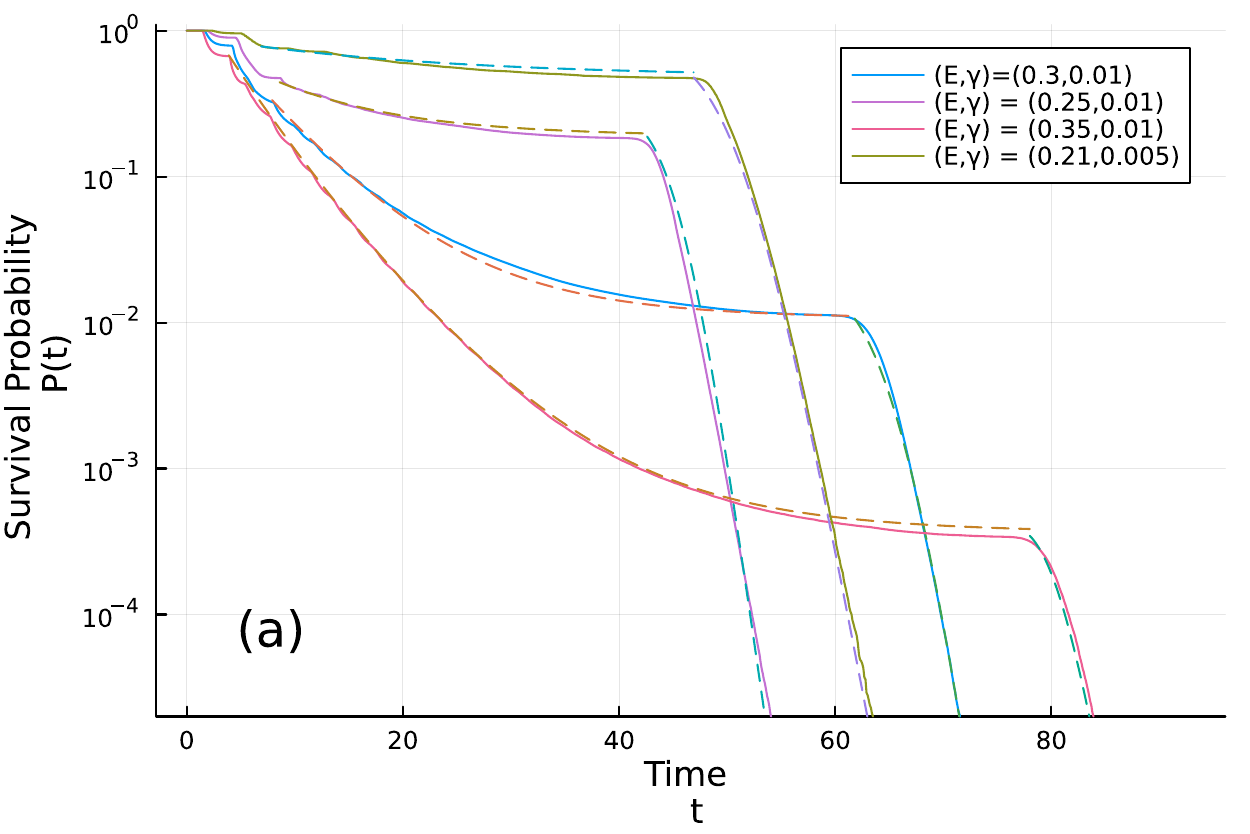}  \includegraphics[width=0.48\linewidth]{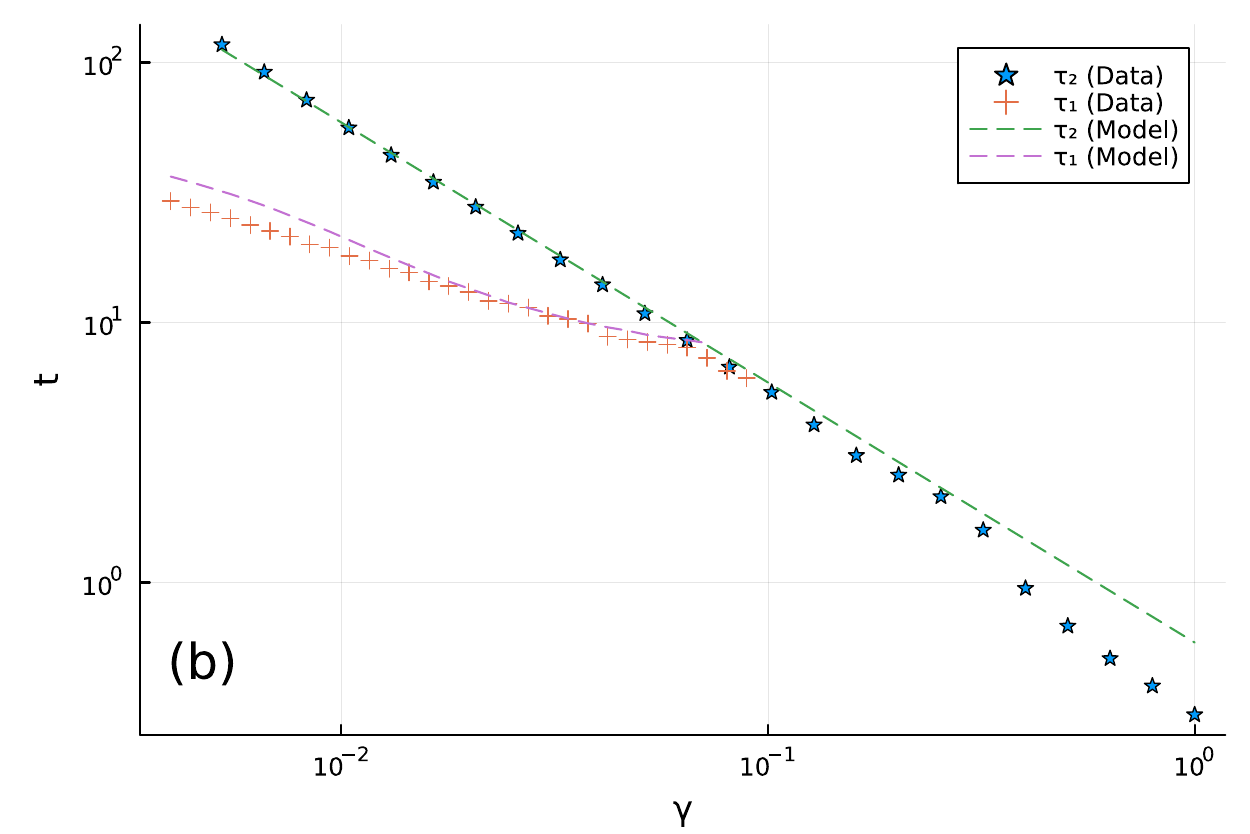}
    \caption{Comparison between theoretical model and numerical simulations of the survival probability~$P(t)$. \textbf{(a)}: Comparison of Eq.~\eqref{eqn:surv_prob_model} (dashed lines) to the simulation data (solid lines) for several choices of $(E,\gamma)$ for $t > \tau_0$. $N_0 = 10^7$ initial conditions were used for each curve. \textbf{(b)}: Comparison of the predicted occurrences (dashed lines) for the transition times $\tau_1$ and $\tau_2$ (see legend) with their observed occurrences (symbols) in the simulations (for $E_0 = 0.3$).
    For $\tau_1$, the observed occurrence is the first time Eq.~\eqref{eqn:tau1} was satisfied in the simulation, while the predicted occurrence comes from numerically solving Eq.~\eqref{eqn:surv_prob_model} with the relevant parameters. For $\tau_2$, the observed occurrence is the first time at which the energy of a non-escaping orbit dropped below $E_c$, and the predicted occurrence $\tau_2 \propto 1/\gamma$ comes from Eq.~\eqref{eq:tau2}. For increasing $\gamma$, the volume of escaping orbits goes to 0 and $\tau_1$ cannot be determined. 
    }
    \label{fig:model_data_comparison}
\end{figure*}

\subsubsection{Conditionally-invariant measure}

As a numerical test of our theory, we turn back to the Hénon–Heiles system. 
Neglecting the variation in the energy of trajectories -- see Fig.~\ref{fig:energy_decay} --, our expectation is that the distribution of long-living trajectories at time $t$ will be well-approximated by the c-measure $\mu_c ^{E}$, with $ E =  E (t)$ given by Eq.~\eqref{eqn:energy_exp_decay}. To verify this, we initialise an ensemble of initial conditions in the dissipative system at an energy $E_2 > E_1$ and, at the time $t^{\star}$ that Eq.~(\ref{eqn:energy_exp_decay}) predicts  $E(t^\star) = E_1$, we plot the position $(x,y)$ of the next crossing with the Poincaré surface (see Appendix~\ref{app:initial_conditions}) and $\mu_c^{E_1}$ (computed using the conservative dynamics).  

The results of one such numerical experiment are reported in Fig.~\ref{fig:um_comparison}(a). 
Our choice of $E_1,E_2,$ corresponds to the two c-measures shown previously in Fig.~\ref{fig:contour_two_manifolds}(b) and $\gamma$ ensures that the time $t^\star$ for the average energy to reach $E_1$ was $\tau_0 \ll t^\star < \tau_2$, allowing enough time for the initially uniform ensemble to relax. We find that the distribution of points in the dissipative system indeed mostly coincides with the c-measure $\mu_c ^{E = E_1}$, which provides a better description than the non-dissipative measure $\mu_c^{E=E2}$ (shown in Fig.~\ref{fig:contour_two_manifolds}(b)). Still, the points of the dissipative system show a more noisy distribution and often do not overlap with (the support of) $\mu_c ^{E = E_1}$. Under Eq.~\eqref{eqn:c-measure_model} the explanation for this is that these points experience a different average decay in energy, implying that their energy at $t \approx t^{\star}$ is far enough from $E_1$ that they are distributed according to different measures (i.e., the disagreement is due to the approximation of neglecting fluctuations in $E(t)$). To test this expectation, in Fig~\ref{fig:um_comparison}(b) we further restrict the distribution to points whose energy is $E(t) \approx E_1$, effectively re-weighting the density $\rho$ towards $E_1$. The reduction in the dispersion of points is clear: the remaining points almost exactly coincide with the support of the measure $\mu_c ^{E = E_1}$. 
The agreement is expected to hold up to a finite spatial scale, which becomes increasingly small as the time for the distributions to relax increases (e.g., by increasing the initial energy $E_2$ or reducing the dissipation rate $\gamma$). For sufficient relaxation time, we posit that the sub-ensemble with energy $E$ converges \emph{exactly} to the c-measure $\mu_c ^E$.

\subsubsection{Survival Probability}
Interpreting the macroscopic dynamics of the dissipative system in terms of the conservative system also allows us to propose a model for the survival probability of escaping orbits. Our c-measure theory predicts that an instantaneous escape rate $\kappa_E$ given by Eq.~(\ref{eqn:kappa_cmeasure_truetime}) applies for the sub-ensemble at each given energy $E(t)=E$ at time $t$. To compute the leading effect of the varying $\kappa_E$ in time, we neglect the dispersion in orbit energies in our system and assume that most trajectories experience the same dynamics (with associated escape rate $\kappa_0$) at any given time $t$. The escape rate, as we have shown in Fig.~\ref{fig:kappa-E-dependence}, depends intimately on the energy, which in turn decays over time.  By combining these information, we can approximate the decay in survival probability $P(t)$ for the entire dissipative ensemble.

The first step is to compute the time-dependent escape rate of the dissipative system, $\kappa_\gamma(t)$, by introducing the exponential decay of the energy as a function of time in Eq.~\eqref{eqn:energy_exp_decay} into our parametric model for $\kappa_0(E)$ (see Fig.~\ref{fig:kappa-E-dependence}), so that 
\begin{equation}
    \kappa_\gamma(t) = \kappa_0(E(t)).
\end{equation}
The expression for the approximate survival probability $P_{\gamma}(t)$ of escaping orbits is then given as
\begin{equation}
    \label{eqn:Pt_escape_generic}
    P_{\gamma}(t) \sim e^{-\kappa_\gamma t},
\end{equation}
which is similar to the exponential decay of hyperbolic scattering in the Hamiltonian system, although now with a time-dependent escape rate $\kappa_\gamma(t)$ as opposed to a fixed rate $\kappa_0(E)$. This description applies to escaping trajectories only, and thus to times $t < \tau_2$. The behavior of $P(t)$ for $t>\tau_2$ is exponential, dominated by trajectories settling towards the attractor, and thus our theoretical model for the survival probability $P(t)$ is
\begin{widetext}
\begin{equation} \label{eqn:surv_prob_model}
    P_{\gamma}(t) = 
    \begin{cases}
        \text{irregular} & t \in [0,\tau_0]\\
        f_{\text{ne}} + f_{e}\cdot\exp\left(-\kappa_\gamma(t)\cdot(t-\tau_0)\right) & t \in [\tau_0,\tau_2]\\ 
        f_{\text{ne}}\cdot \exp\left(-g(t,\gamma)\cdot (t-\tau_2)\right) & t > \tau_2
    \end{cases}
\end{equation}
\end{widetext}
where $\kappa_\gamma = \kappa_0(E(t))$ is the escape rate at the (average) energy at time $t$.
The term $f_{ne}$ refers to the fraction of trajectories in the initial ensemble that remain trapped, which at fixed $E_0$ increases with $\gamma$ (resp. decreases with increasing $E_0$ at fixed $\gamma$.) Likewise, $f_e$ refers to the fraction of orbits that escape at a time $t > \tau_0$. We have that $f_{ne} + f_{e} = P(\tau_0)$. The function $g(t,\gamma)$ is introduced because the observed transition in $P(t)$ between the escape-dominated and settling-dominated decay regimes is smooth. We thus require $g(t,\gamma)$ to be such that $g(t)\ge 0$ for $t\ge 0,\gamma > 0$ and to increase monotonically with $t$ towards a constant $L$ that depends on the dissipation strength $\gamma$.  In practice, we found that a sigmoid function of the following form provides a satisfying fit
\begin{equation}
    g(t,\gamma) = \frac{L}{( 1+\exp(-b(t-\tau_2))) ^2}.
\end{equation}
Finally, we evaluate Eq.~(\ref{eqn:surv_prob_model}) using the parametric fits of $\kappa_0 (E)$ -- in Fig.~\ref{fig:kappa-E-dependence} -- and $E(t)$ -- in Fig.~\ref{fig:energy_decay}.

\par Direct comparison of simulation data for select parameter values with the predictions from Eq.~\eqref{eqn:surv_prob_model} is shown in Fig.~\ref{fig:model_data_comparison}(a).  In general, our model shows a satisfying agreement with the data, despite relying on the approximation of homogeneous energy decay. 
Fig.~\ref{fig:model_data_comparison}(b) shows the comparison between the observed occurrences of the timepoints $\tau_1$ and $\tau_2$ with their predicted occurrences under our model while varying $\gamma$. 
An improved agreement would be obtained by avoiding the approximation of homogeneous energy decay and better taking into account properties of the energy density $\rho(E(t))$. As an example, this could take the form of a mixture model for $P(t)$ involving multiple sub-ensembles, each with their own energy decay exponent $\mu$. 
These results confirm that the range $\tau_1 \le t \le \tau_2$ of dissipative scattering in which our probabilistic description applies becomes arbitrarily large with reducing $\gamma$.

\section{Discussion and Conclusion}\label{sec:conclusion}

We have presented a generally-applicable model for the study of dissipative chaotic scattering and illustrated its validity with numerical studies of the Hénon–Heiles system. This model proposes that long-living trajectories in the dissipative system can be described, for an intermediate but arbitrarily large time interval $t<\tau_2$, by conditionally-invariant measures $\mu_c$ of the non-dissipative system. Our description includes the dependence of the measures and time interval on dissipation rate~$\gamma$, initial energy $E_0$, and time~$t$. The temporal evolution of ensembles of trajectories in the dissipative system can then be understood by considering weighted sub-ensembles of the conservative system, with each sub-ensemble at a given energy $E(t)$ described by the corresponding c-measure. This theoretical model has shown to be effective at explaining transient chaos phenomena like the survival probability $P(t)$, which can be described using time-dependent escape rates, and the distributions of trajectories in the phase space. 

An interesting observation made in our experiments involving the survival probability is that the theoretically derived curves apply also for times at which trajectories are in energy ranges $E_c < E \le E_{KAM}$ in which the conservative system shows the existence of KAM islands. This is a regime in which there is no c-measure $\mu_c^E$ or escape rate $\kappa$ and therefore our description should not strictly apply. The reason for its unexpected success is that the hyperbolic components of the invariant sets generate an effective escape rate that drives the numerical observations~\cite{Motter2003,Altmann2009,RevModPhys.85.869}. For smaller dissipations $\gamma$, trajectories have time to relax towards the non-hyperbolic components of the saddle in these energy ranges and deviations become relevant. In particular, we expect a concentration of the combined c-measure in Eq.~(\ref{eqn:c-measure_model}) towards non-hyperbolic regions and that $P(t)$ will converge to a constant earlier in $t$ (i.e., when the energy reaches $E_{KAM}$ instead of $E_c$).

Our probabilistic description provides also clear predictions for the fractal properties of scattering systems with dissipation, a problem previously studied through different approaches~\cite{MotterAdilsonE.2013DtcG,Karolyi_2021,Chen2017,Zotos2017,Vilela_2021}. These studies employed a variety of different theoretical and numerical approaches (e.g., based on the uncertainty algorithm~\cite{GrebogiCelso1983FssA}) to evaluate the fractality of basin geometry as a function of the dissipation. Our proposed description is that the evolution of the dissipative system can be viewed as a time-evolving mixture of snapshots of the Hamiltonian system and that a well-defined fractal distribution exists (for arbitrarily small spatial scales) in these systems in suitable limits of $E_0,\gamma,$ and $t$ that allow for a relaxation towards $\mu_c^E$. In more general numerical settings, a cut-off in the scaling underlying fractal analysis (in time or space) is expected due to the combination of $\mu_c^E$s (each with a slightly different fractal support). An interesting question for future work is in which extent our probabilistic description retrieves or reproduces these previous results.

Another area of future work is to extend the probabilistic description for the dynamics of ensembles of initial conditions using an operator-theoretic formalism. Our results showing that the evolution depends mainly on the time-dependent energy density $\rho(E(t))$ suggest that further insights can be obtained through spectral approaches considering Perron-Frobenius (or similar) operators acting on these densities~\cite{FroylandGary2007Doco,Froyland2008,RevModPhys.85.869}.

\appendix

\section{Numerical methods}\label{app:numerics}

\subsection{Integrators}

Integration of differential equations was performed in using the \verb|DifferentialEquations.jl|~\cite{rackauckas2017differentialequations} and \verb|DynamicalSystems.jl|~\cite{Datseris2018} software libraries for the \verb|Julia| programming language. Visualisations were made using the \verb |Plots.jl|~\cite{PlotsJL} library.

For integrating trajectories of the Hamiltonian system, we used an 8th-order symplectic integrator~\cite{kahan1997composition,rackauckas2017differentialequations} with timestep $dt = 0.01$. Conservation of the Hamiltonian was typically at the limits of double precision, even for very long integrations. For the dissipative system, we used an efficient 9th-order Runge-Kutta integrator with adaptive timestepping~\cite{verner2010numerically}. In both cases a maximum integration time of $5000$ was used, which was more than sufficient to observe the escape/settling of all orbits not bound inside KAM islands.  

\subsection{Initial Conditions}\label{app:initial_conditions}

For the purposes of testing our theory, and to faciliate visualisation, we chose to have all points initialised with the same energy. We achieved this by sampling from a Poincaré surface of section that allowed us to specify the initial positions $(x_0 ,y_0)$ and energy $E$ freely and then determine the initial velocities $(\dot{x}_0,\dot{y}_0)$ to match. Specifically, we chose the section corresponding to zero radial velocity ($\dot{r} = 0$) and positive angular velocity ($\dot{\phi} > 0$). The initial velocities $(\dot{x}_0,\dot{y}_0)$ in Cartesian coordinates under this section condition are given by 
\begin{align}
&\dot{x}_0 =  -\frac{y_0}{\sqrt{x_0^2 + y_0^2}}\sqrt{2(E-V(x_0,y_0))}\\
&\dot{y}_0 = \frac{x_0}{\sqrt{x_0^2 + y_0^2}}\sqrt{2(E-V(x_0,y_0))}
\end{align}

\subsection{Escape Condition}\label{app:escape_condition}
In the Hamiltonian setting $(\gamma = 0)$ we used a numerical criterion to determine escape. If the orbit's $(x,y)$ position exceeds a fixed radius $r_{esc} > 1$ then the orbit is deemed to have escaped at this time. The distance from the origin to any of the three saddle points is $1$, so choosing an $r_{esc}$ slightly larger than $1$ ensures that the orbit has actually crossed the threshold determined by the Lyapunov orbit and thus left $\Omega$. Various specific values of $r_{esc}$ were considered, all leading to the same observations.
\medskip
\par In the dissipative setting $(\gamma > 0)$, the Lyapunov orbit criterion cannot be relied upon, so we choose an expedient method to determine the escape condition: we integrate an initial condition until the energy of the orbit drops below the critical energy $E_c$ . An orbit with energy $E < E_c$ cannot leave or re-enter $\Omega$, and so its position at this time will determine its eventual fate. An orbit on the exterior of $\Omega$ will be considered to have escaped, and an orbit on the interior will be considered trapped. Integrating further is unnecessary due to the monotonicity of the energy decay. However, for an escaping orbit, we consider the actual escape time to be the time when the trajectory left $\Omega$ $(r > r_{esc})$, with the apriori knowledge that it will not return. Trajectories that leave $\Omega$ quickly achieve high velocity, and so rapidly fall below $E_c$ on the outside.


\providecommand{\noopsort}[1]{}\providecommand{\singleletter}[1]{#1}%

\end{document}